\documentclass[aps,preprint,superscriptaddress,floatfix,showpacs,showkeys]{revtex4}
\usepackage{graphics}

\begin{document}
\newenvironment{tab}[1]
{\begin{tabular}{|#1|}\hline}
{\hline\end{tabular}}

\title {Interface effects and termination of finite length nanotubes}

\author{N. Stefanakis}
\affiliation{Universit\'e Libre de Bruxelles, C.P. 231
B-1050, Brussels, Belgium
}  
\date{\today}

\begin{abstract}
The objective of the present paper 
is to investigate interface effects in
carbon nanotubes. We use both real and k-space 
tight binding method. We study in detail the effect 
of wrapping vector on the electronic properties.
We analyze the effect of the curvature in closed 
nanotube to the electronic properties. The finite length 
of the nanotube affects the electronic properties.  
\end{abstract}
\pacs{}
\maketitle

\section{Introduction}

A carbon nanotube is obtained when the honeycomb
graphene sheet is wrapped into a seamless cylinder \cite{iijima}. 
The electronic properties
are sensitive on its geometrical structure.
Depending on the orientation of the hexagons with 
respect to the nanotube axis, they can be classified
to 'armchair', 'zigzag', and 'chiral'.
The single wall nanotubes 
(SWNT), which consist of a single graphite layer, 
have diameter around $1-2$nm and they are proposed 
as
ideal components for the future nano devices. 

Scanning tunneling microscopy and spectroscopy on individual 
SWNT has been used recently to provide atomically resolved images 
and to allow the determination of the nanotube electronic properties 
as a function of the wrapping vector and diameter \cite{wildoer}.

The electronic properties of hybrid structures of carbon nanotubes 
have already been analyzed theoretically within the Hubbard model 
(see in our previous work \cite{stefan3}). We pointed out there that 
the finite length of nanotubes affects the properties of the 
hybrid structure. 

In this paper we calculate within the real and $k$-space tight binding 
method the electronic properties of carbon nanotubes. We 
examine the effect of the chirality on the electronic properties. 
In addition we study termination of nanotubes i.e. the effect of 
the curvature at the cups of nanotube to the local density of states.
We point out the differences between the two approaches. The real space 
method has the advantage that it permits the treatment of finite size 
samples.

In the following we describe the methods in Sec II.
In Sec III we present the results using the k-space and 
real space tight binding method. We discuss the effect of 
chirality, and the termination of the 
nanotubes. 
In the last section we present the conclusions.

\section{method}
\subsection{k-space tight binding method}
Tight binding theory in k-space has been used to calculate 
the electronic properties of nanotubes. 
The energy dispersion 
relation for nanotube is found from the two dimensional 
dispersion relation for the $\pi$ bands 
of graphite by eliminating one of the two components of the 
wave vector $k$ according to the periodic boundary 
conditions in the circumferential direction \cite{saito}
\begin{equation}
C_h\dot k=2\pi m, 
\end{equation}
where $C_h=na_1+ma_2$ is the chiral vector, $a_1, a_2$ are the 
unit cell basis vectors of graphite, and $n,m$ are integers. 
For example the energy dispersion 
relation for armchair nanotube is
\begin{equation}
E_q^a(k)=\pm t\{1\pm 4 \cos(q\pi/n)\cos(ka/2)+4 \cos^2(ka/2)\}^{1/2}.
\end{equation} 
For zig-zag nanotubes the corresponding dispersion relation reads 
\begin{equation}
E_q^z(k)=\pm t\{1\pm 4 \cos(q\pi/n)\cos(\sqrt{3}ka/2)+4 \cos^2(q\pi/n)\}^{1/2}.
\end{equation} 
Using the one dimensional dispersion relations the density of states (DOS)
is calculated using the following formulas \cite{economou}
\begin{equation}
\rho(E)=\pm\frac{1}{\pi}im\{G(l,l;E)\},
\end{equation} 
where the diagonal matrix element of the Green function $G(l,l;z)$ in 
one dimension is 
\begin{equation}
G(l,l;z)=\frac{1}{2\pi}\int_{-\pi}^{\pi}d\phi \frac{1}{z-E(\phi)}.
\end{equation} 

\subsection{real space tight binding method}
Within the second approach we describe the
SWNT
by exact diagonalizations of the real space tight binding 
Hamiltonian in a honeycomb lattice\cite{stefan,stefan1,stefan2,stefan3}.
The corresponding Hamiltonian is written as
\begin{equation}
H = -t\sum_{<i,j>\sigma}c_{i\sigma}^{\dagger}c_{j\sigma}
+\mu \sum_{i\sigma} n_{i\sigma}
,~~~\label{bdgH}
\end{equation}
where $i,j$ are sites indices and the angle brackets indicate that the
hopping is only to nearest neighbors,
$n_{i\sigma}=c_{i\sigma}^{\dagger}c_{i\sigma}$ is the electron number
operator in site $i$, $\mu$ is the chemical potential.
Within tight binding method we solve the following eigenvalue 
problem 
\begin{equation}
\hat{\xi}u_n(r_i)=\epsilon_n u_n(r_i)
,~~~\label{bdg}
\end{equation}
where
\begin{equation}
\hat{\xi}u_n(r_i)=-t\sum_{\hat{\delta}}
u_n(r_i+\hat{\delta})+\mu u_n(r_i),~~~\label{bdgxi}
\end{equation}
and we obtain the eigenvectors $u(r_i)$ and
the eigenenergies $\epsilon_n$ self consistently.
Then the local density of states (LDOS) at the $i$th site 
is calculated by
\begin{equation}
\rho_i(E)=-2\sum_{n}
\left [ |u_n(r_i)|^2 f^{'}(E-\epsilon_n)
\right ]
,~~~\label{bdgdos}
\end{equation}
where the factor $2$ comes from the twofold spin degeneracy,
$f^{'}$ is the derivative of the Fermi function,
\begin{equation}
f(\epsilon)=\frac{1}{\exp(\epsilon/k_B T) + 1}.
\end{equation}
For numerical stability reasons and in order to present more 
realistic results we set the temperature $T$ to a non zero value.

SWNT is formed by rolling the honeycomb sheet into a
cylinder and using the appropriate boundary conditions to 
describe armchair and zigzag structures.

\section{results}
\begin{figure}
\centering\scalebox{0.8}{\includegraphics{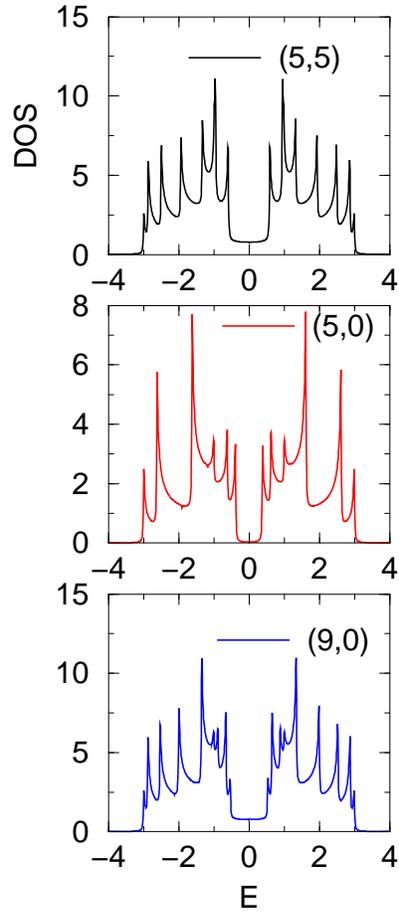}}
\caption{The DOS for several metallic and semiconducting 
nanotubes using the tight binding k-space approach.
}
\label{green.fig}
\end{figure}

We present first the well known results using the 
tight binding method in reciprocal space. We note 
that this method is restricted to description of 
nanotubes of infinite length.
In Fig. \ref{green.fig} we show the DOS for several nanotube structures.
In order to analyze the density of states 
we have to take into account the dispersion relation of nanotubes. 
The dispersion relation of nanotube consists of $2n$ curves for the 
valence and conduction bands respectively. For 
armchair nanotube the valence and conduction bands 
cross at the Fermi level. Therefore the armchair nanotube is 
metallic as indicated by the finite density of states 
at $E=0$ in Fig. \ref{green.fig}. We also observe 
singularities at the band edges which correspond to the extrema 
of $E(k)$ relations. For zig-zag nanotube ($n,0$) when $n$ is not 
multiply of $3$ the dispersion relation curves do not 
cross at the Fermi level. As a consequence the nanotube 
is semiconducting with an energy gap at the Fermi energy as 
seen in Fig. \ref{green.fig} for the ($5,0$) nanotube.
An exception occurs when $n$ is a multiply of $3$. 
For example in the $(9,0)$ nanotube which 
is shown at the bottom of the figure, we note that 
although its geometry corresponds to a zig-zag nanotube, it 
is actually metallic since analysis of the energy dispersion 
relation shows that the valence and conduction bands cross 
each other at the Fermi level at the $\Gamma$ point ($k=0$).  

\begin{figure}
\centering\scalebox{0.6}{\includegraphics{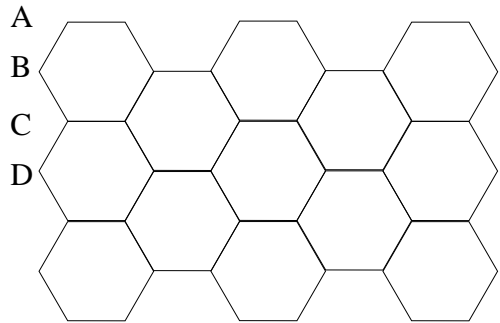}}
\caption{The graphite lattice close to the corner.
}
\label{grafitelattice.fig}
\end{figure}

\begin{figure}
\centering\scalebox{0.6}{\includegraphics{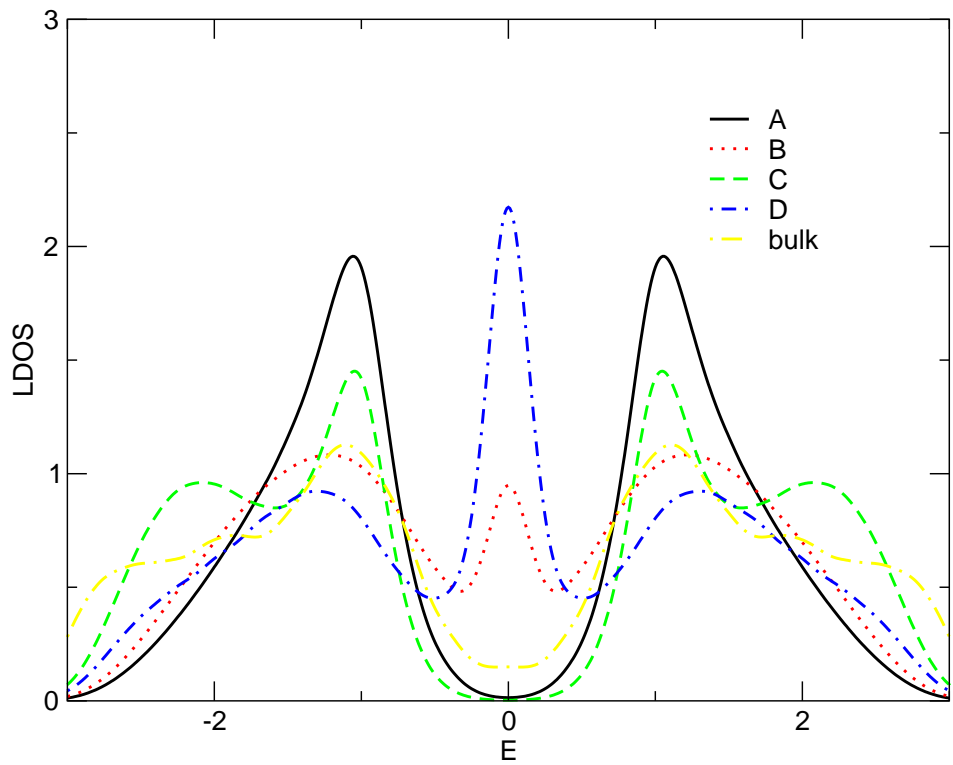}}
\caption{The LDOS for points A,B,C,D seen in Fig. \ref{grafitelattice.fig}
and the bulk LDOS. 
}
\label{grafite.fig}
\end{figure}

\begin{figure}
\centering\scalebox{0.6}{\includegraphics{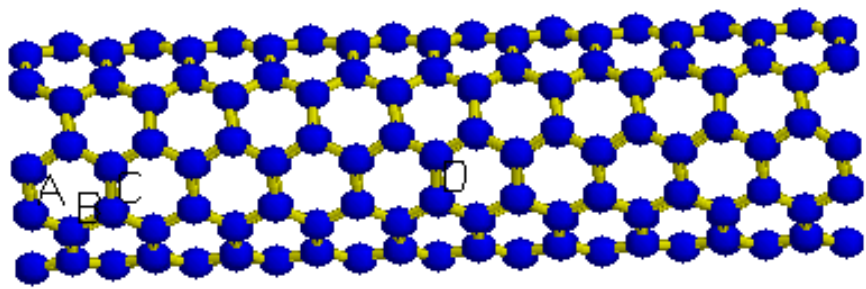}}
\caption{The open armchair $(5,5)$ nanotube composed of 21 layers.
}
\label{55.fig}
\end{figure}

\begin{figure}
\centering\scalebox{0.6}{\includegraphics{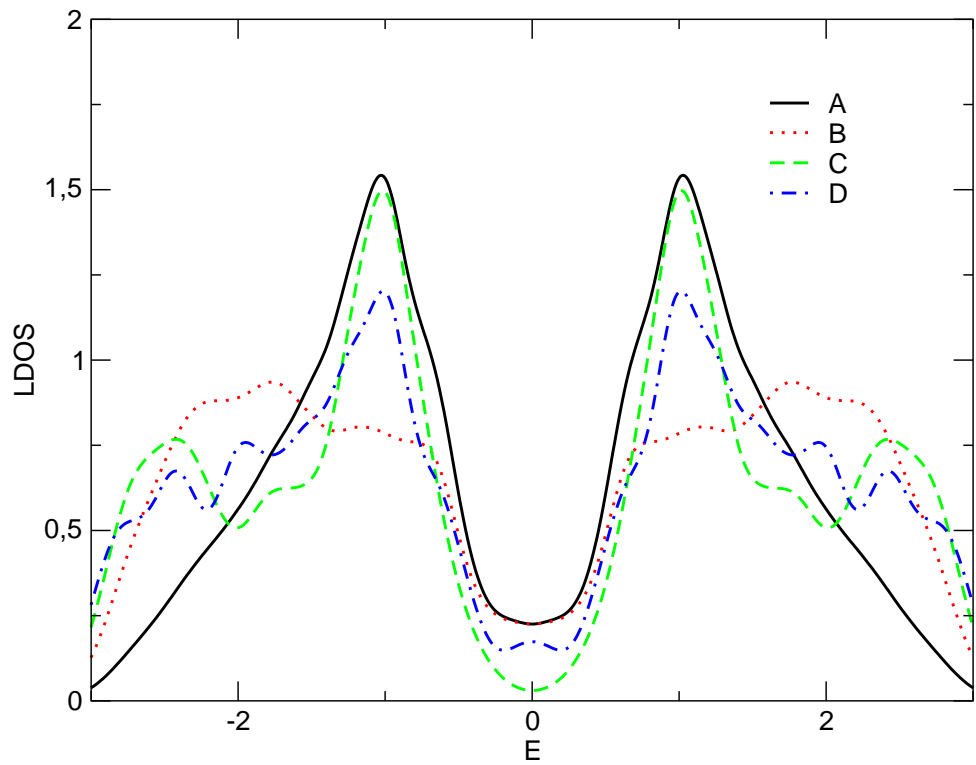}}
\caption{The LDOS for points A,B,C,D of the armchair $(5,5)$ nanotube 
seen in Fig. \ref{55.fig}.
}
\label{LDOS55.fig}
\end{figure}

We present now the results for the real space tight binding 
method. Contrary to the previous case which is restricted to 
infinite length 
nanotubes, this approach permits the description of structures that 
are of finite length. Since the modern nanodevices require 
finite length components this approach is more advanced 
compared to the $k$ space approach.
The LDOS close to the
boundary
for the two dimensional graphite lattice which is presented 
in Fig. \ref{grafitelattice.fig}
shows strong deviation from the bulk value 
(see Fig. \ref{grafite.fig}). It is well known that the 
bulk graphite is zero gap semiconductor which is not the 
case for graphene nanosamples close to the interface since there 
are curves with finite DOS at the Fermi level.
We study then the armchair $(5,5)$ nanotube 
seen in Fig. \ref{55.fig}. The LDOS shown in 
Fig. \ref{LDOS55.fig} shows finite states in the Fermi level 
which indicate that the SWNT is metallic. This agrees with 
the previous $k$-space approach. However we find that the LDOS 
is strongly modulated close to the boundary (see sites A,B,C in
Fig. \ref{LDOS55.fig}) from the bulk D site. Even-more at specific 
sites e.g. C the LDOS approaches zero which indicates insulating 
behavior. An other characteristic is the presence of bands with 
one-dimensional Van-Hove singularities at the band edges. 
Due to the finite temperature that we use the overall line 
shape is smooth. 

\begin{figure}
\centering\scalebox{0.6}{\includegraphics{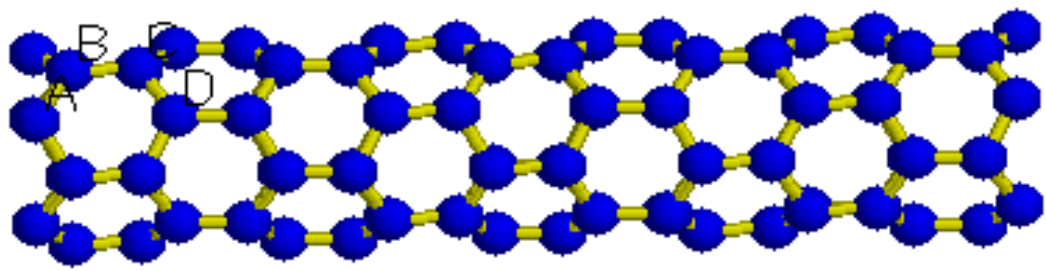}}
\caption{The open
zigzag $(5,0)$ nanotube composed of 20 layers.
}
\label{50.fig}
\end{figure}

\begin{figure}
\centering\scalebox{0.6}{\includegraphics{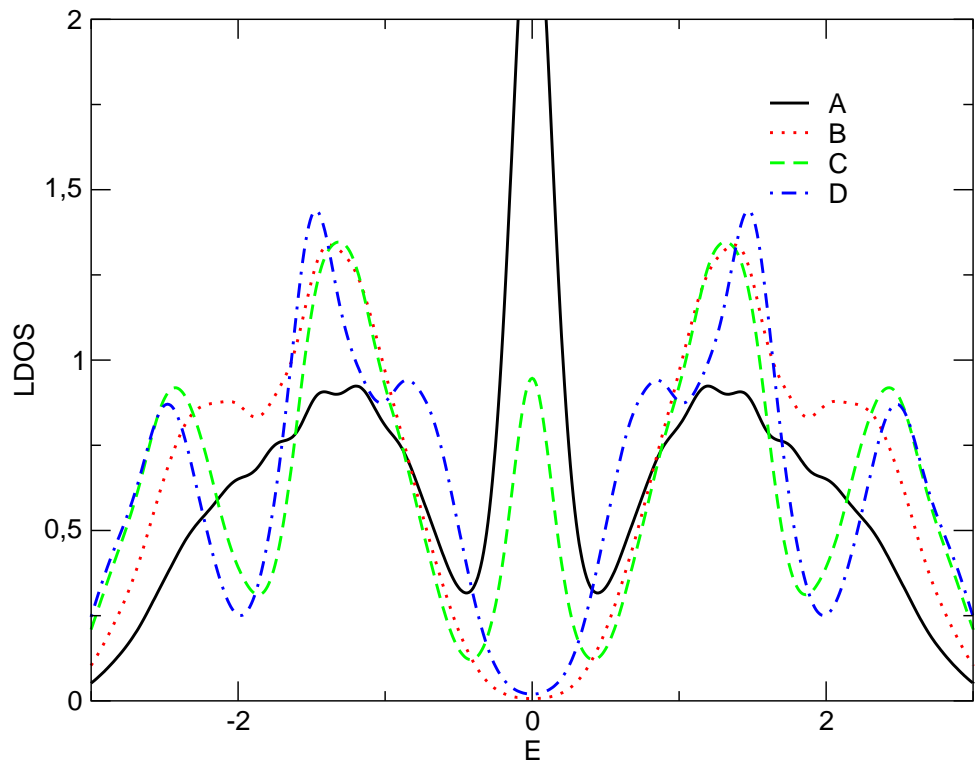}}
\caption{The LDOS for points A,B,C,D of a zig-zag $(5,0)$ nanotube.
}
\label{LDOS50.fig}
\end{figure}

We would like to describe other nanotubes like the zig-zag 
which is shown in Fig. \ref{50.fig}. This nanotube is expected 
to be semiconducting. We confirm that for the bulk. However close to the 
surface a peak appears at specific sites (see Fig. \ref{LDOS50.fig}). 
This is opposite to the $k$-space method. 

\begin{figure}
\centering\scalebox{0.6}{\includegraphics{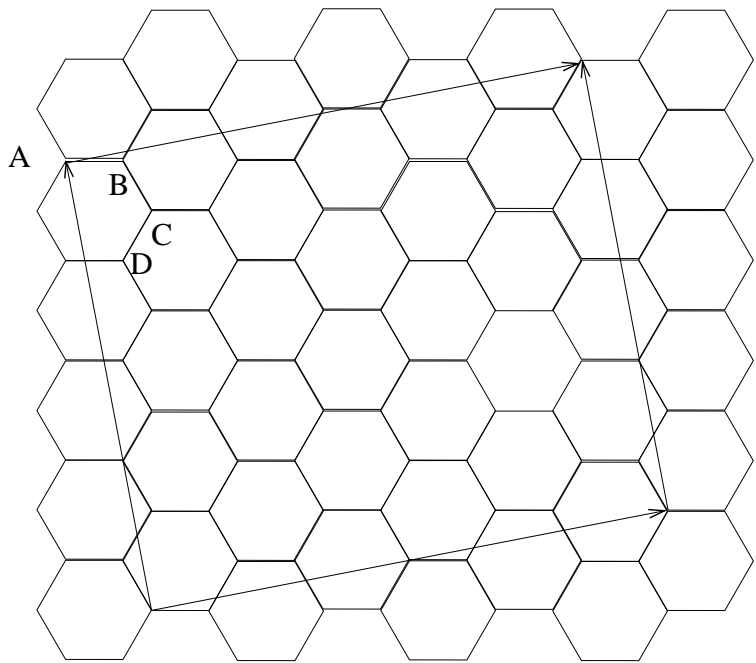}}
\caption{The open
chiral $(4,2)$ nanotube.
}
\label{chiral42.fig}
\end{figure}

\begin{figure}
\centering\scalebox{0.6}{\includegraphics{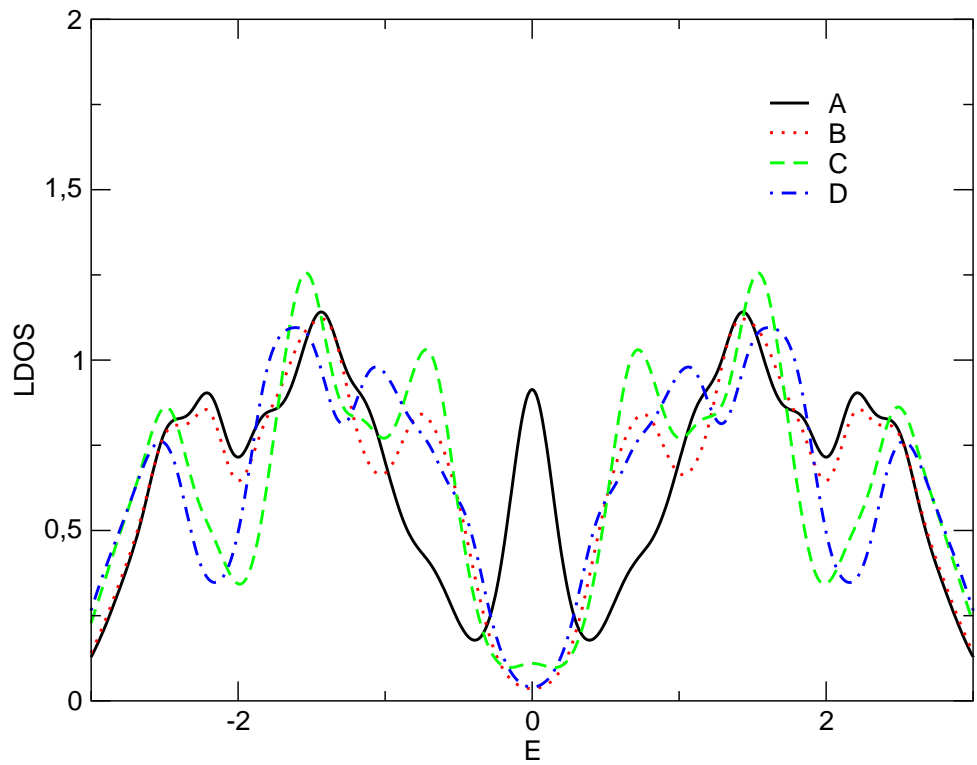}}
\caption{The LDOS for points A,B,C,D of a chiral $(4,2)$ nanotube.
}
\label{LDOS42.fig}
\end{figure}

We also describe the chiral nanotube
which is shown in Fig. \ref{chiral42.fig}. This nanotube is
semiconducting since $n-m$ is not a multiply of $3$. 
The magnitude of the semiconducting gap depends on the 
chiral vector and is different than the other semiconducting 
structures that we studied e.g. $(5,0)$.
However close to the 
surface a peak appears at specific sites (see Fig. \ref{LDOS42.fig}). 

\begin{figure}
\centering\scalebox{0.6}{\includegraphics{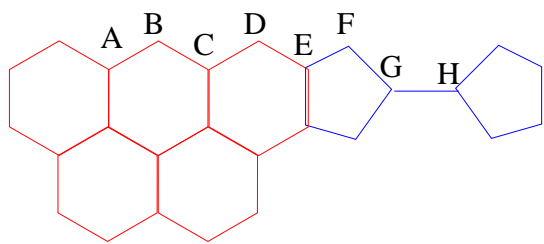}}
\caption{The closing of an armchair $(5,5)$ nanotube.
}
\label{closing55lattice.fig}
\end{figure}

\begin{figure}
\centering\scalebox{0.6}{\includegraphics{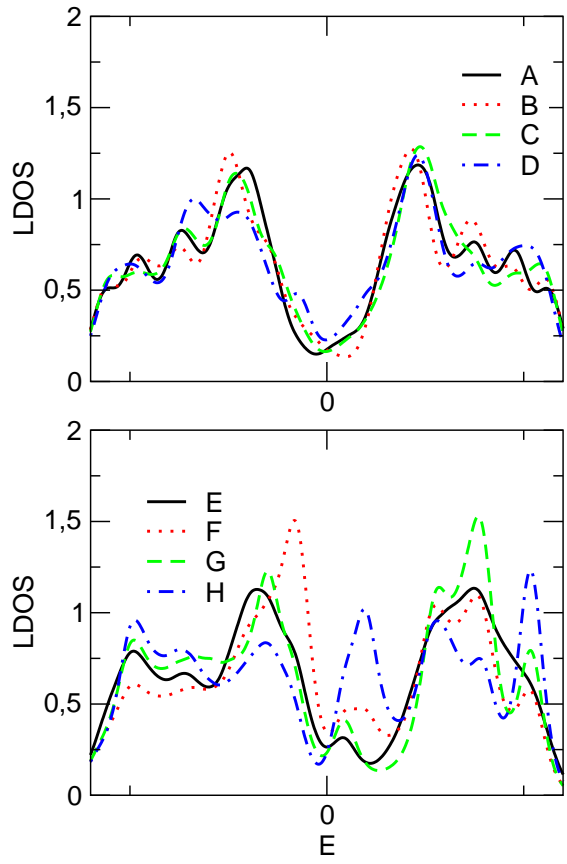}}
\caption{The LDOS for the points A,B,C,D,E,F,G,H that are seen in the 
previous figure closing the end of an armchair $(5,5)$ nanotube.
}
\label{closing55.fig}
\end{figure}

\begin{figure}
\centering\scalebox{0.6}{\includegraphics{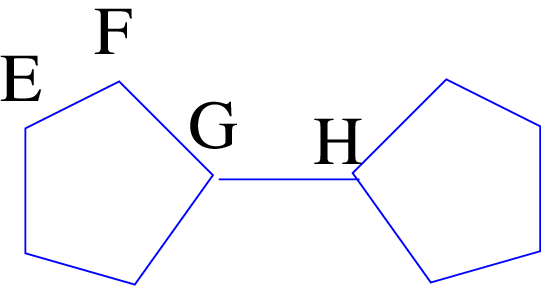}}
\caption{The geometry of a $(5,5)$ nanocone.
}
\label{cup55lattice.fig}
\end{figure}

\begin{figure}
\centering\scalebox{0.6}{\includegraphics{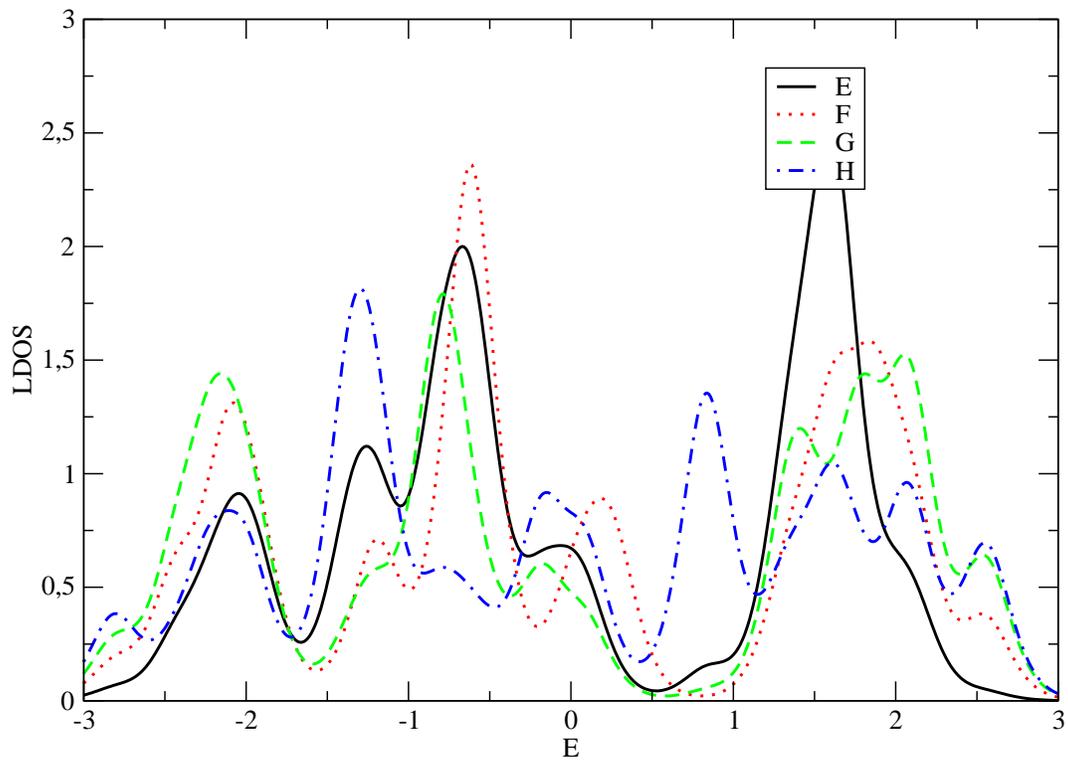}}
\caption{The LDOS for the points E,F,G,H that are seen in the 
figure \ref{cup55lattice.fig} at the $(5,5)$ nanocone.
}
\label{cup55.fig}
\end{figure}

We discuss the termination of nanotubes. We specifically show that 
a geometrical and topological effect like the introduction of 
pentagons in a hexagonal lattice affects not only the curvature 
of the nanotube but also its electronic properties. We show in 
Fig. \ref{closing55lattice.fig} the termination of an armchair nanotube
using a cup which consists of a hemisphere of a fullerene. 
The cup contains six pentagons and an appropriate number of 
hexagons. The LDOS close to the end of the nanotube
is deviating from the metallic behavior as
seen in Fig. \ref{closing55.fig}. As we can observe 
a peak is developing close to Fermi energy which
corresponds to localized states that are formed due to the deformation 
of the lattice by the introduction of the pentagons. 
 In order to investigate the effect of the cup to the nanotube we 
study then the electronic structure of the isolated nanocone. This 
is formed by placing six pentagons and appropriate number of hexagons 
such as a cone structure is formed (see in Fig. \ref{cup55lattice.fig}).
This cone nanostructure closes exactly the $(5,5)$ nanotube and 
constitutes a half fullerene molecule.
As is seen the bound states still exist. However due to the 
drastic reduction of the number of atoms the produced nanostructure 
has density of states which resembles that of a molecule with 
discrete bound levels (see Fig. \ref{cup55.fig}).

\begin{figure}
\centering\scalebox{0.6}{\includegraphics{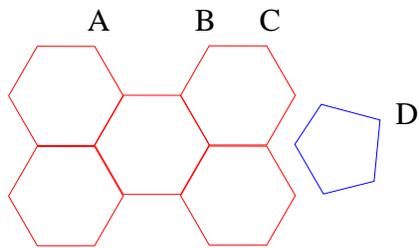}}
\caption{The closing of a zig-zag $(5,0)$ nanotube.
}
\label{closing50lattice.fig}
\end{figure}

\begin{figure}
\centering\scalebox{0.6}{\includegraphics{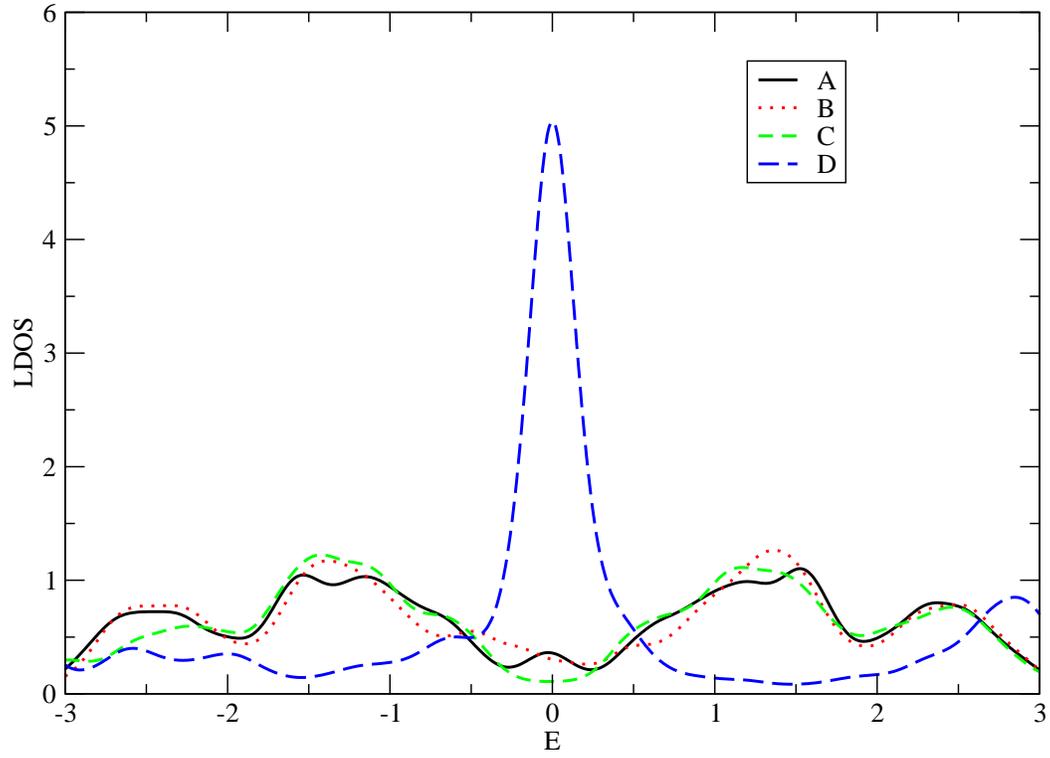}}
\caption{The LDOS for the points A,B,C,D that are seen in the 
previous figure closing the end of a zig-zag $(5,0)$ nanotube.
}
\label{closing50.fig}
\end{figure}

\begin{figure}
\centering\scalebox{0.6}{\includegraphics{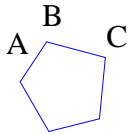}}
\caption{The geometry of a $(5,0)$ nanocone.
}
\label{cup50lattice.fig}
\end{figure}

\begin{figure}
\centering\scalebox{0.6}{\includegraphics{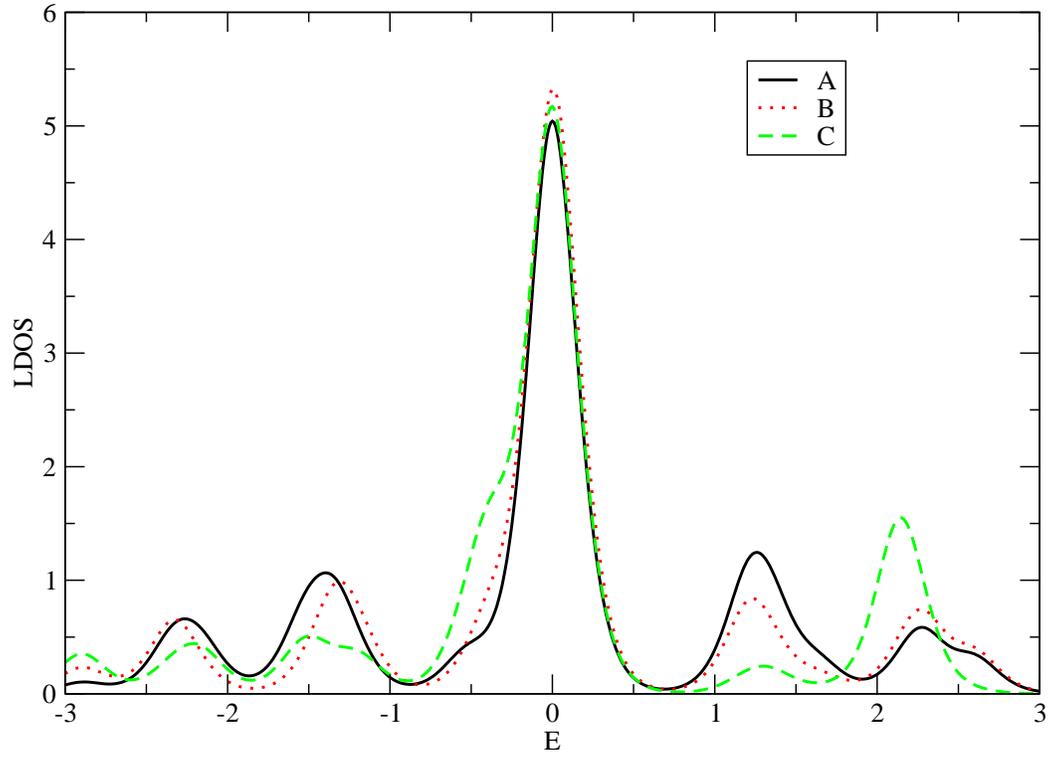}}
\caption{The LDOS for the points A,B,C that are seen in the 
figure \ref{cup50lattice.fig} at the $(5,0)$ nanocone.
}
\label{cup50.fig}
\end{figure}

Using the same procedure is possible to terminate a zig-zag nanotube.
We show in 
Fig. \ref{closing50lattice.fig} the termination of a zig-zag nanotube
using appropriate number of pentagons.
The LDOS close to the end of the nanotube
is 
seen in Fig. \ref{closing50.fig}. Here the peak which corresponds 
to 
the localized states around the pentagons exists at the Fermi energy. 
 In order to investigate the effect of the cup to the $(5,0)$ nanotube we 
study then the electronic structure of the isolated nanocone. This 
is formed by placing six pentagons and appropriate number of hexagons 
such as a cone structure is formed (see in Fig. \ref{cup50lattice.fig}).
This cone nanostructure closes exactly the $(5,0)$ nanotube.
As is seen the bound states still exist. However due to the 
drastic reduction of the number of atoms the produced nanostructure 
has density of states which resembles that of a molecule with 
discrete bound levels (see Fig. \ref{cup50.fig}).

\section{conclusions}
We have studied the electronic properties of finite length 
SWNT with in the 
tight binding method self consistently. 
This method provides advantages compared to the $k$-space method, 
namely that it permits the study of finite length nanotubes, 
which are expected to become the building blocks of the future nanodevices.
The results indicate that for finite length nanotubes the 
LDOS is strongly modified close to the boundary layers of the 
nanotube. Moreover the local density of states changes 
considerably around the pentagons that terminate a nanotube 
where additional bound states are formed.

We have to make a distinction between the topological defects 
which we have studied in the present work and which do 
affect the electronic and transport properties of nanotubes and the 
mechanical deformations which result from bending twisting, and  
compressing the nanotube. The later in the absence of 
topological defects have minor effect to the electronic 
properties. 

When studying the electronic properties of nanotubes the 
$sp^2$ hybridization of the atomic orbitals which has been used 
here, produces realistic results provided that the 
diameter of the nanotube is large enough. Concerning purely 
mechanical properties the $sp^3$ hybridization would produce 
more realistic results. 

\section{acknowledgments}
This work was supported in part from a stipend from ULB.

\bibliographystyle{prsty}


\newpage
\vspace{20cm}

\newpage
\vspace{20cm}

\end{document}